\documentclass[manuscript]{aastex}
\usepackage{amssymb}

\usepackage{epsfig}
\usepackage{psfrag}
\usepackage{amsfonts}          
\usepackage{amsmath}           
\usepackage{xspace}
\usepackage{rotating}
\usepackage{natbib}
\usepackage{booktabs}
\usepackage{color}

\providecommand{\eg}{e.g.\xspace}
\def\bs{\begin{subequations}}
\def\es{\end{subequations}}
\def\be{\begin{equation}}
\def\ee{\end{equation}}
\def\bml{\begin{multline}}
\def\eml{\end{multline}}
\def\beq{\begin{eqnarray}}
\def\eeq{\end{eqnarray}}
\def\bd{\begin{description}}
\def\ed{\end{description}}
\def\bn{\begin{enumerate}}
\def\en{\end{enumerate}}
\def\bi{\begin{itemize}}
\def\ei{\end{itemize}}
\def\brem{\begin{remark}}
\def\erem{\end{remark}}
\def\bp{\begin{proof}}
\def\ep{\end{proof}}

\renewcommand\r{\right}

\newcommand{\dd}[2]{\partial_{#1}{#2}}

\newcommand{\f}[2]{\frac{#1}{#2}}

\renewcommand{\vec}[1]{\mbox{\boldmath $ #1$}}
\renewcommand{\v}{\vec}

\newcommand{\B}{\vec B}
\renewcommand{\u}{\vec u}
\newcommand{\m}{\vec v}
\newcommand{\ru}{\hat{\vec r}}
\renewcommand{\r}{\vec r}
\newcommand{\kk}{\hat{\vec k}}
\newcommand{\n}{\nabla}
\renewcommand{\.}{\cdot}
\newcommand{\x}{\times}
\newcommand{\rhobar}{\bar{\rho}}
\newcommand{\Tbar}{\bar{T}}
\newcommand{\Pbar}{\bar{P}}
\newcommand{\nubar}{\bar{\nu}}
\newcommand{\kapbar}{\bar{\kappa}}
\newcommand{\Nrho}{N_\rho}
\newcommand{\pol}{v}
\newcommand{\tor}{w}
\newcommand{\polB}{h}
\newcommand{\torB}{g}

\newcommand{\D}{\mathcal{D}}
\newcommand{\E}{\mathcal{E}}
\newcommand{\M}{\mathcal{M}}

\newcommand{\Dl}{\D_l}
\newcommand{\El}{\E_l}
\newcommand{\Ml}{\M_l}
\newcommand{\Lmax}{{L_\text{max}}}

\newcommand{\R}{\mathrm{R}}
\renewcommand{\Pr}{\mathrm{Pr}}
\newcommand{\Pm}{\mathrm{Pm}}

\newcommand{\Rm}{\mathrm{Rm}}
\newcommand{\Ro}{\mathrm{Ro}}
\newcommand{\Lo}{\mathrm{Lo}}
\newcommand{\Nu}{\mathrm{Nu}}

\makeatother

\shorttitle{}
\shortauthors{Simitev et al}

\allowdisplaybreaks

\newlength{\shortwidth}
\newlength{\textwidtha}

\usepackage{xr}
\newcommand{\hide}[1]{}

\newcommand{\vtwors}[1]{#1}
\newcommand{\revrs}[2]{#2}
\newcommand{\revrsa}[1]{#1}

\usepackage{soul}

\begin{document}


\title{Dynamo Effects Near Transition from Solar to Anti-Solar
  Differential Rotation}

\author{Radostin D.~Simitev\altaffilmark{1,2,3}, Alexander G.~Kosovichev\altaffilmark{3,4}, and Friedrich H.~Busse\altaffilmark{2,5}}

\altaffiltext{1}{School of Mathematics and Statistics, University of
  Glasgow -- Glasgow G12 8QW, UK}
\altaffiltext{2}{Department of Earth and Space Sciences, University of
  California, Los Angeles -- Los Angeles, CA 90095, USA}
\altaffiltext{3}{NASA Ames Research Center -- Moffett Field, CA 94035, USA}
\altaffiltext{4}{New Jersey Institute of Technology -- Newark, NJ 07103, USA}
\altaffiltext{5}{Institute of Physics, University of Bayreuth --
  Bayreuth 95440,  Germany}

\begin{abstract}
Numerical MHD simulations play increasingly important role for
understanding mechanisms of stellar magnetism. We present simulations
of convection and dynamos in \revrsa{density-stratified} rotating spherical
fluid shells. \revrsa{We employ a new 3D simulation code for 
the solution of a physically consistent anelastic model of the process
with a minimum number of parameters.} 
The reported dynamo simulations extend into a ``buoyancy-dominated''
regime where the buoyancy forcing is dominant while the Coriolis force
is no longer balanced by pressure gradients and strong anti-solar
differential rotation develops as a result. We find that 
the self-generated magnetic fields, despite being relatively weak, are
able to reverse the direction of differential rotation from anti-solar
to solar-like. We also find that convection flows in this regime are
significantly stronger in the polar regions than in the equatorial
region, leading to non-oscillatory dipole-dominated dynamo solutions,
and to concentration of magnetic field in the polar regions. We observe
that convection has \revrsa{different morphology} in the inner and
at the outer part of the convection zone simultaneously such that
organized geostrophic convection columns are hidden below a
near-surface layer of well-mixed highly-chaotic convection. \revrsa{While we
focus the attention on the buoyancy-dominated regime, we also demonstrate
that conical differential rotation profiles and persistent regular
dynamo oscillations can be obtained in the parameter space of the
rotation-dominated regime even within this minimal model.}
\end{abstract}


\newpage

\section{Introduction}


Recent progress in observational and computational capabilities  have
led to substantial advances in our understanding of the origins of stellar
and planetary magnetism and variability. The key role is played by
dynamo processes driven by convection in the presence of rotation. This
process is particularly complex in the case of turbulent convection
in stellar and planetary envelopes. The interaction of convection,
magnetic field  and rotation results in variations of the rotation
rate with depth and latitude, called \emph{differential rotation}, and
also in large-scale meridional flows. The differential rotation is a
crucial part of dynamo mechanisms. It has been measured quite
accurately for the solar surface  by tracking motion of various
features and also through analysis of the Doppler shift of spectral
lines. Moreover, helioseismology data from the SOHO and SDO space missions
and from the ground-based network GONG have provided measurements of the internal
rotation  \citep[e.g.][]{Schou1998} and of the meridional circulation of the
Sun \citep{Zhao2013}.  

Recently, high-precision spectroscopic and photometric observations have
enabled measurements of the differential rotation on other stars. The
most prominent feature of the solar differential rotation is that the
equatorial zone rotates faster than higher latitude regions.  For
other stars, this type of differential rotation is called
\textit{solar-like} rotation. When the equator rotates slower than the
polar regions then such rotational profile is called
\textit{anti-solar}. 
\revrs{q5}{Anti-solar differential rotation has been observed by
Doppler imaging techniques on several K-giant stars \citep{Strassmeier2003,Kovari2015}.}
Recent analysis of the high-precision light curves from the Kepler
mission for a sample of 50 G-type stars by \citet{Reinhold2015}
found 21--34 stars with the solar-like differential rotation and 5--10
stars with the anti-solar rotation.  \revrs{q5}{The latter work
awaits confirmation from independent studies.}   

One of the first theories of non-linear convection in rotating shells
developed in the Boussinesq approximation  by \citet{Busse1970}, \revrsa{also \citep{Busse1973}},
demonstrated that the dynamical effects of rotation on convection are
a primary mechanism of stellar differential rotation. This theory
revealed that the basic properties of differential rotation primarily
depend on the supercritical value of the Rayleigh number which
measures the strength 
of the buoyancy forces and determines the magnitude of convection
motions. For a relatively small supercritical Rayleigh numbers 
convection develops mostly in the equatorial region in the form of
convective rolls (also known as "banana cells") oriented along the rotation
axis. Angular momentum transport by these cells causes solar-like
differential rotation. This regime of weakly supercritical Rayleigh
numbers corresponds to small Rossby numbers defined as the ratio of the
rms  convective velocity to the mean rotational velocity. This regime
is called \textit{rotationally-dominated}. Early numerical simulations in
the Boussinesq approximation by \citet{Gilman1976} confirmed these
results, and also found that at high supercritical Rayleigh numbers when
convection develops at all latitudes differential rotation may
become anti-solar. This regime is characterized by large (typically greater than
1) Rossby numbers \revrsa{and is called
  \textit{buoyancy-dominated}.}. 
 While the rotationally-dominated regime is likely to occur
in the deep convection zone where convective velocities are small,
\citet{Gilman1979} noticed that on the Sun a typical supergranulation
turn-over time is much shorter than the rotation period, and thus the
convection is in the bouyancy-dominated regime. This suggests
that on the Sun a combination of the two regimes takes place, resulting
in a complicated differential rotation profile.

Indeed, the differential rotation \revrsa{profile of the Sun}
determined by helioseismology \citep[e.g.][]{Schou1998} turned out to
be quite different from the theoretical predictions. 
\revrsa{It is characterised by nearly radial orientation of
  angular velocity at midlatitudes 
(the so-called ``conical profile''). The angular velocity also decreases 
monotonically from the equator to the poles by about 30\%.} In
addition, helioseismology inferences reveal two narrow rotational
shear layers at the boundaries of the convection zone: the so-called
\emph{tachocline} at the bottom and a \emph{near-surface shear layer} at the top.

Substantial efforts have been made to reproduce the observed solar rotation
in \revrs{q12}{\revrs{q13}{three-dimensional numerical simulations.} These efforts are
reviewed by \citet{Miesch2005} and more recently by \citet{Brun2014a}. 
Historically, simulations have been successful in reproducing the
decrease of angular velocity from equator to poles. However, it is a
major discrepancy that most simulations tended to feature angular
velocity contours parallel to the rotation axis (the so-called
``cylindrical profiles''), rather than conical profiles.
The profile of the differential rotation is governed by the zonal
component of the vorticity equation and its analysis suggests that 
cylindrical profiles may be avoided if (a) baroclinic forcing
with non-vanishing latitudinal entropy gradient exists, or
if (b) sufficiently strong Reynolds stresses or (c) Lorentz  forces
develop \citep[e.g.][]{Miesch2005,Miesch2006}. Pursuing alternative (a),
\cite{Miesch2006} imposed a latitudinal gradient of entropy as a
bottom boundary condition in global convection simulations. They were
so successful in finding conical differential rotation profiles that
many groups have subsequently adopted this model as standard
\citep[e.g.][]{Browning2006,Nelson2013,Fan2014}. While there is physical
justification for a non-vanishing latitudinal entropy gradient,
this model assumption must be regarded with caution as the imposed
latitudinal variations in entropy have not been generated
self-consistently by the convective motions in the simulations. Large
amplitude Reynolds stresses required to realize alternative (b) can be achieved 
by driving convection more strongly. Because most early studies argued
that the solar magnetic field cannot exert a substantial control 
over differential rotation, alternative (b) has been explored most often.
 A recent work in this direction is} the attempt of 
\citet{Guerrero2013} to find a regime close to the real solar
rotation  by varying  the gradient of the background specific entropy
and the  frame rotation rate in a numerical anelastic model. These
variations correspond to an increasing effective Rayleigh number
\revrsa{and thus to a more strongly driven convection}.  However, the simulations did not find an intermediate
regime with a ``conical'' profile. Instead they surprisingly 
showed that the transition between the solar and anti-solar rotational
profiles is rather sharp. \revrsa{Most recently} similar results have been
reported by \citet{Gastine2013,Gastine2014,Kapyla2014,Mabuchi2015},
and \citet{Karak2015}.  

\revrsa{Alternative (c), namely that dynamo effects are crucial in
shaping the profile of differential rotation, is much less
explored. A significant number of dynamo simulations have been
published previously, but most of them aimed to reproduce and explain
solar magnetic features and activity cycles,
\citep[e.g.][]{Browning2006,Ghizaru2010,Nelson2013}. Only recently few
studies, of which we mention \citep{Fan2014,Mabuchi2015}, have
appeared that aim to investigate whether the solar magnetic field
plays an active role in shaping the solar differential rotation
profile. These studies report evidence in support of this
hypothesis. This is hardly surprising as} it is well-established that
the main effect of  self-sustained magnetic field on convection is to
suppress differential rotation 
\citep[e.g.][]{GroteBusse2001,Busse2003,Simitev2005a}. \revrs{q7}{This
finding is confirmed by \cite{AUBERT2005} and \cite{Yadav2013}
who also provide scaling laws for this effect.} \revrs{q2}{However, the
effects of dynamo action on convection}  \revrs{q3}{in the
buoyancy-dominated regime are less well explored. Studies
include the afore mentioned papers by
\citep{Fan2014,Karak2015,Mabuchi2015} who report a number of similar
results but their models are rather different from each other and in
all cases also include factors that contribute to baroclinic forcing
such that hypothesis (c) is not addressed in isolation.} 

\revrs{q2}{In this context our paper has several goals. \revrs{q3}{We wish to} study the
effects of self-generated magnetic fields on the convective flows of a
density-stratified fluid in rotating spherical shell using a minimal
self-consistent model. We wish to focus the attention on dynamos near
the transition to buoyancy-dominated convection but we also report
selected results in the rotation-dominated regime. To this end, we employ
the so-called Lantz-Braginsky anelastic approximation
\citep{BraginskyRoberts1995,LantzFan1999,Jones2011}. It has the
advantages that the dynamics of the system depends on a minimal number 
of non-dimensional parameters while \emph{ad-hoc} parametrizations of 
physical effects that may be used to better ``fit'' observations are
excluded. We present a newly implemented numerical simulation code for
the solution of the Lantz-Braginsky anelastic equations, as
well as code validation results based on published benchmark solutions
against four other independently-developed codes \citep{Jones2011}.
Since the dynamo processes operating in the Sun and stars are still
subjects to controversies with various competing proposals, we believe
that it is important that our results are readily reproducible by other groups.} 

In Section \ref{sec.2} we introduce the mathematical model based on
the anelastic approximation, and discuss the numerical method and
diagnostic output. In Section \ref{sec.3} we present the benchmark
validation results. In Section \ref{sec.4}  we discuss the properties
of dynamos in the buoyancy-dominated regime of convection.
\revrs{q4}{\revrs{q12}{In Section \ref{sec.49} we demonstrate that conical
differential rotation profiles and persistent regular dynamo
oscillations can be obtained in the rotation-dominated regime.}}
\revrsa{In Section \ref{sec.5}, we present} a summary of our main
results \revrs{q2}{\revrs{q3}{and compare them to related recent
 studies}}. We also outline topics for future research.

\section{Mathematical model and numerical method}
\label{sec.2}

We consider an electrically conducting, perfect gas confined to a spherical shell. The shell rotates with a fixed angular
velocity $\Omega \kk$ about the vertical axis and an entropy contrast
$\Delta S$ is imposed between its inner and outer surfaces. 

\subsection{Anelastic governing equations}

Assuming a gravity field proportional to $1/r^2$, a hydrostatic
polytropic reference state exists of the form  
\begin{gather}
\rhobar = \rho_c\zeta^n, \quad \Tbar=T_c\zeta, \quad \Pbar = P_c
\zeta^{n+1}, \quad
\zeta= c_0+c_1 d/r,
\end{gather}
with parameters $c_0=(2\zeta_o-\eta-1)/(1-\eta)$,
$c_1=(1+\eta)(1-\zeta_o)/(1-\eta)^2$, $\zeta_o=(\eta+1)/(\eta
\exp(\Nrho/n)+1)$.
The parameters $\rho_c$, $P_c$ and $T_c$ are reference values of
density, pressure and temperature at the middle of the shell. The
gas polytropic index $n$, the density scale height number, 
$N_\rho$, and the radius ratio, $\eta$, are defined below. 
Convection and magnetic field generation are described by the 
equations of continuity, momentum, energy and magnetic flux. In the
anelastic approximation \citep{Gough1969,BraginskyRoberts1995,LantzFan1999,Jones2011} these equations
take the form   
\bs
\label{gov}
\begin{gather}
\label{gov.01}
\n\.\rhobar\u =0, \quad \n\.\B =0, \\
\label{gov.02}
\partial_t \u + (\n\x\u)\x\u 
=-\n\Pi -\tau(\kk\x\u)+\f{\R}{\Pr}\f{S}{r^{2}}\ru
+ \v F_\nu + \f{1}{\rhobar} (\n\x\B)\x\B, \\
\label{gov.03}
\partial_t S + \u\.\n S 
= \f{1}{\Pr \rhobar\Tbar} \n\.\kapbar\rhobar\Tbar \n S
  + \f{c_1 \Pr}{\R \Tbar}\left(Q_\nu+ \f{1}{\Pm\rhobar} Q_j\right), 
\\
\label{gov.04}
\partial_t \B = \n\x(\u\x\B)+\Pm^{-1} \n^2 \B,
\end{gather}
\es
where $\u$ is the velocity, $\B$ is the magnetic flux density, $S$
is the entropy and $\nabla \Pi$ includes all terms that can be
written as gradients. The viscous force ($\v F_\nu$), and the viscous
($Q_\nu$) and Joule ($Q_j$) heating are defined in terms of the
deviatoric stress tensor  ($\hat S_{ij}$) 
\begin{gather}
\label{tens}
\hat S_{ij}=2\nubar\rhobar(e_{ij}-e_{kk}\delta_{ij}/3), \quad e_{ij}=(\partial_i u_j +\partial_j u_i)/2,
\nonumber
\end{gather}
\begin{gather}
\label{not}
\v F_\nu = \frac{\rho_c}{\rhobar}\n\.\v {\hat S}, \quad Q_\nu=\v{\hat S}:\v e,  \quad Q_j=(\n\x\B)^2, 
\end{gather}
where the double-dots symbol (\textbf{:}) denotes the Frobenius inner product.
We assume that the viscosity $\nu$ and the entropy diffusivity
$\kappa$ are constant throughout the shell.
The governing equations are parametrized using
the thickness of the  shell $d=r_o-r_i$ as a unit of length,
$d^2/\nu$ as a unit of time, $\Delta S$ as a 
unit of entropy, $\nu \sqrt{\mu_0\rho_c}/d$ as a unit of magnetic
induction, $\rho_c$ as a unit of density and $T_c$ as a unit of
temperature. Here, $r_i$ and $r_o$ are the inner and the outer radius,
$\lambda$ and $\mu_0$ are the magnetic diffusivity and permeability,
respectively. The system is then characterized by eight
non-dimensional parameters: the radius ratio $\eta=r_i/r_o$, the
polytropic index $n$, the density scale number
$N_\rho=\ln\big(\rhobar(r_i)/\rhobar(r_o)\big)$,  the Rayleigh number $\R={c_1 T_c d^2 \Delta S}/{(\nu
  \kappa)}$, the thermal Prandtl number $\Pr={\nu}/{\kappa}$, the 
magnetic Prandtl number $\Pm={\nu}/{\lambda}$, and the Coriolis
number $\tau = {2\Omega d^2}/{\nu}$.   

Since the mass flux $\m\equiv\rhobar \u$, and the magnetic flux
density $\vec B$ are solenoidal vector fields, we
employ a decomposition in poloidal and toroidal components,  
\bs
\label{poltordef}
\begin{gather}
\rhobar \vec u = \nabla \times ( \nabla \times \ru r\pol) + \nabla \times
\ru r^2 \tor, \\
\vec B = \nabla \times  ( \nabla \times \ru \polB) + \nabla \times
\ru \torB,
\end{gather}
\es
where $\ru$ is the radial unit vector, $r$ is the length of the position vector $\vec r$, $\pol$,
$\tor$, $\polB$ and $\torB$ are the poloidal and toroidal scalars of
the momentum and magnetic field, respectively.
Equations~\eqref{gov.01} are then satisfied automatically. Scalar
equations for $v$ and $w$ are obtained, and effective pressure
gradients are eliminated by taking $\ru\cdot\n\x\n\x$ and
$\ru\cdot\n\x$ of Equation~\eqref{gov.02}. Similarly, equations for
$h$ and $g$ are obtained by taking $\ru\cdot \n\x$ and $\ru\cdot$ of
Equation~\eqref{gov.04}. \vtwors{Spectral projections of the resulting
poloidal-toroidal equations are presented in Appendix \ref{appendix},
see also Section \ref{nummeth}.}

\subsection{Boundary conditions}

Equations \eqref{gov} must be supplemented by boundary conditions.
For the simulation results presented in this work the following
boundary conditions are used. The inner and the outer surfaces of the
shell are assumed as stress-free, impenetrable boundaries for the flow
\begin{gather}
\pol = 0,
\quad
\partial_r^2 \pol - \f{\rhobar'}{\rhobar r}\dd{r}{} (r\pol) = 0,
\quad
\partial_r \tor - \f{\rhobar'}{\rhobar} \tor = 0, \quad
\text{at}\quad r=r_i,r_o.
\end{gather}
\revrsa{In several cases discussed in Section \ref{sec.49} we use
no-slip condition at the inner boundary,
\begin{gather}
\pol = 0,
\quad
\dd{r}{} \pol = 0,
\quad
\tor=0, \quad \text{at}\quad r=r_i.
\end{gather}}
A fixed contrast of the entropy is imposed between the inner and the
outer surface
\begin{gather}
S=1 \text{ at } r=r_i, \quad S=0 \text{ at } r=r_o.
\end{gather}
The boundary conditions for the magnetic field are derived from the
assumption of an electrically insulating external regions. The
poloidal function $\polB$ is then matched to a function $\polB^{(e)}$, which
describes an external  potential field,
\begin{gather}
\torB =0, \quad  \polB-\polB^{(e)} = 0, \quad \partial_r
(\polB-\polB^{(e)})=0 \quad
\text{at}\quad r=r_i,r_o.
\end{gather}
We remark that our code presented below allows various other choices
of boundary conditions to be made for all of the dynamical variables.

\subsection{Numerical method}
\label{nummeth}
To perform the numerical simulations of this study, we have extended
our Boussinesq code
\citep{TilgnerBusse1997,Busse2003,Simitev2005a,Simitev2006b,Simitev2008a,Simitev2009,Simitev2012a}
to solve the anelastic equations described in Section \ref{sec.2}.
Despite similarities with the Boussinesq code, this is a major
modification both in terms of the mathematical model and the numerical
code. For the numerical solution of the problem we have adapted the
pseudo-spectral method described by \cite{Tilgner1999}. 
The scalar unknowns $\pol$, $\tor$, $\polB$, $\torB$ and $S$, are
expanded in Chebychev polynomials $T_p$ in the radial direction $r$, and
in spherical harmonics in the angular directions $(\theta,\varphi)$ e.g.,
\begin{equation}
\label{decomp}
\pol = \sum\limits_{l=1}^{N_l} \sum\limits_{m=-l}^{l}
\sum\limits_{p=0}^{N_r} V_{lp}^m (t)\, T_p\Big(x(r)\Big)\, P_l^m ( \cos \theta )\, \exp(im \varphi),
\end{equation}
where $P_l^m$ denotes the associated Legendre functions, 
$x(r)=2(r-r_i)-1$, and $N_l$ and $N_r$ are truncation parameters.
A system of equations for the coefficients in these expansions is
obtained by a combination of a Galerkin spectral projection of the
governing equations in the angular directions and a collocation
constraint in radius. \vtwors{This system is presented in Appendix
\ref{appendix}.} 

Computation of nonlinear terms in spectral space is expensive,
so nonlinear products and the Coriolis term are computed in the physical
space and then projected into the spectral space at every time
step. A standard 3/2-dealiasing in $\theta$ and $\varphi$ is used
at this stage. 
A hybrid of a Crank-Nicolson scheme for the diffusion terms and a
second order Adams-Bashforth scheme for the nonlinear terms is used
for integration in time.

\begin{figure}
\epsfig{file=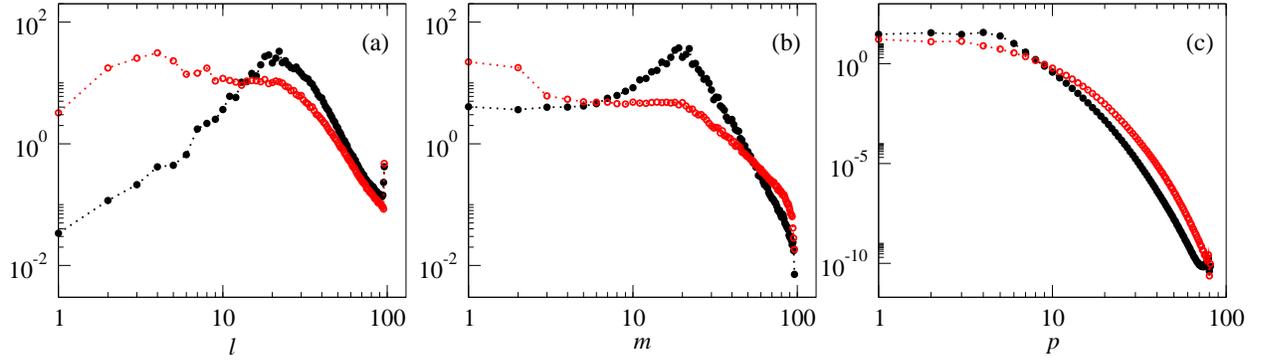,width=\textwidth,clip=}
\caption{
Time-averaged power spectra of kinetic (full circles) and magnetic
(empty circles) energy as a function of (a) the harmonic degree $l$,
(b) the harmonic order $m$ and (c) the Chebyshev polynomial degree $p$ in the case 
$\eta=0.65$, $\R=6\times10^6$, $\Pr=1$, $\Pm=2$, $\tau=2000$, $n=2$,
  $N_\rho=3$. 
}
\label{FIG0001.eps}
\end{figure}

\revrs{q8}{Calculations are considered adequately resolved when the
spectral power of kinetic and magnetic energy drops by more than two
orders of magnitude from the spectral maximum to the cut-off wavelength as
suggested \eg~by \cite{Christensen1999}.}  
A range of numerical resolutions has been employed in this study varying
from ($N_r=61$, $N_l=96$)  in less demanding cases to ($N_r=121$,
$N_l=144$) in more strongly stratified or turbulent
runs. Correspondingly, the physical gridpoints on which
non-linear terms are evaluated have been varied up to $N_r=121$,
$N_\theta=216$, $N_\varphi=437$.  \revrs{q8}{We find that this
provides adequate resolution as demonstrated in Figure
\ref{FIG0001.eps} for a typical dynamo solution.}

\revrs{q1}{
The pseudo-spectral approach described above is the most common method
for solving the fundamental equations of convection-driven flow and
electromagnetic induction within a rotating spherical shell filled
with an electrically conducting fluid. The approach was pioneered by 
\cite{Glatzmaier1984} and with appropriate modifications it has been
widely used by various groups for modelling convection-driven geo-,
Solar, and planetary dynamos. A number of codes based on similar
principles have been developed that differ mainly in details such as
time stepping methods and treatment of radial dependence with
finite differencing and Chebyshev decomposition in $r$ being two
popular choices. Early versions of some codes were derived directly
from the code of \cite{Glatzmaier1984}, including the anelastic ASH
code extensively used for Solar simulations \citep{Clune1999} and the
Boussinesq MAG code used for geodynamo simulations
\citep{OlsonChristensenGlatzmaier1999}. Other codes including ours were
developed independently 
\citep[e.g.][]{Jones1995,Tilgner1999,Hollerbach2000}. While it is not feasible to
provide here a comprehensive list of existing numerical codes and discuss the numerous
variations in actual implementation, we refer to a series of
benchmarking papers \citep{Christensen2001,Jones2011,Marti2014,Jackson2014} and to the reviews
\citep{Miesch2005,WichtTilgner2010} for overview of codes commonly
used in the solar context and in the geo-/planetary context,
respectively. These references also include discussions of other
essentially different numerical approaches of solution. Recent studies
related to ours \citep{Fan2014,Karak2015,Mabuchi2015,Hotta2015} use
finite difference methods to solve comparable but different sets of
equations. Local methods have better parallel efficiency but also
inferior accuracy \citep{Tilgner1999}. They also have difficulties
with imposing global boundary conditions for the magnetic field,
e.g.~all of the above dynamo models use the unphysical radial
condition for the magnetic field on the outer surface. They also encounter 
difficulties in treating spherical geometries,
e.g.~\cite{Fan2014} and \cite{Karak2015} consider wedges and not 
spherical shells. Our code has been developed and optimized
independently  over a number of years and a large database of
Boussinesq results is available for comparison. The current anelastic
version of our code is new and perhaps unique among other spectral
codes, with the exception of the ASH code, in allowing for a radial
dependence of the viscosity and the thermal and magnetic
diffusivities. However, the latter facilities have not been used in the present analysis.   
}

\subsection{Diagnostic output quantities} 

We characterise convection and dynamo solutions 
by their kinetic and magnetic energy and heat transport given by a
Nusselt number.  
The energies can be conveniently split into poloidal and toroidal
components, mean and fluctuating components and further into 
equatorially-symmetric and equatorially-antisymmetric components. We thus
obtain a good characterisation of the scales of the convective
flow and the multipole structure of dynamos. The mean and fluctuating
toroidal and poloidal components of the total kinetic energy
$E_\text{kin}$ are defined as  
\bs
\label{engs}
\begin{eqnarray}
\label{7}
\bar E_p = \langle \big(\nabla \times ( \nabla \bar \pol \times \vec r )
\big)^2/(2\rhobar)  \rangle,
& 
\bar E_t = \langle \big(\nabla r \bar\tor \times \vec r \big)^2
/(2\rhobar)  \rangle, \\
\label{8}
\check E_p =  \langle \big(\nabla \times ( \nabla \check \pol
\times \vec r)  \big)^2/(2\rhobar) \rangle, &
\check E_t = \langle \big( \nabla r \check \tor \times\vec r \big)^2
/(2\rhobar) \rangle,
\end{eqnarray}
\es
where  angular brackets $\langle \rangle$ denote averages over the
spherical volume of the shell, overlaid lines denote axisymmetric
parts and overlaid check marks denote non-axisymmetric parts of a
scalar field. The total magnetic energy $E_\text{magn}$ can be split
in a similar way with components defined as in Equations \eqref{engs}
but with $\polB$ and $\torB$ replacing $\pol$ and $\tor$ and 
without the factor $\rhobar^{-1}$ within the angular brackets. The
total energies are, of course, the sum of all components. The Nusselt
number is defined as the ratio between the values of the luminosity of
the convective state $\mathrm{L_{conv}}$ and the luminosity of the
basic conduction state $\mathrm{L_{basic}}$, 
$$
\Nu=\frac{\mathrm{L_{conv}}}{\mathrm{L_{basic}}},\qquad 
\mathrm{L_{conv}}=-\int_{\partial V(r)}\kappa\rhobar\Tbar (\partial_r S)
\r^2 \sin\theta d\theta d\varphi, 
\quad
\mathrm{L_{basic}}=\frac{4\pi n c_1 \rhobar(r)^n}{\exp(\Nrho)-1},
$$
with the integral taken over a spherical surface $\partial V(r)$ at
radius $r$. Apart from
quantifying the 
heat transport of convection, the value of the Nusselt number serves as a
convenient proxy for the super-criticality of the convective regime.

Other diagnostic quantities used below are the non-dimensional
magnetic Reynolds number, 
$\Rm= \Pm \sqrt{2 E_\text{kin}}$,  
Rossby number, 
$\Ro= \dfrac{2}{\tau}\sqrt{2 E_\text{kin}}$,  
and Lorentz number 
$\Lo=\dfrac{2}{\tau}\sqrt{2 E_\text{magn}}$. 

\subsection{Benchmarking and validation}
\label{sec.3}
\begin{figure}
\epsfig{file=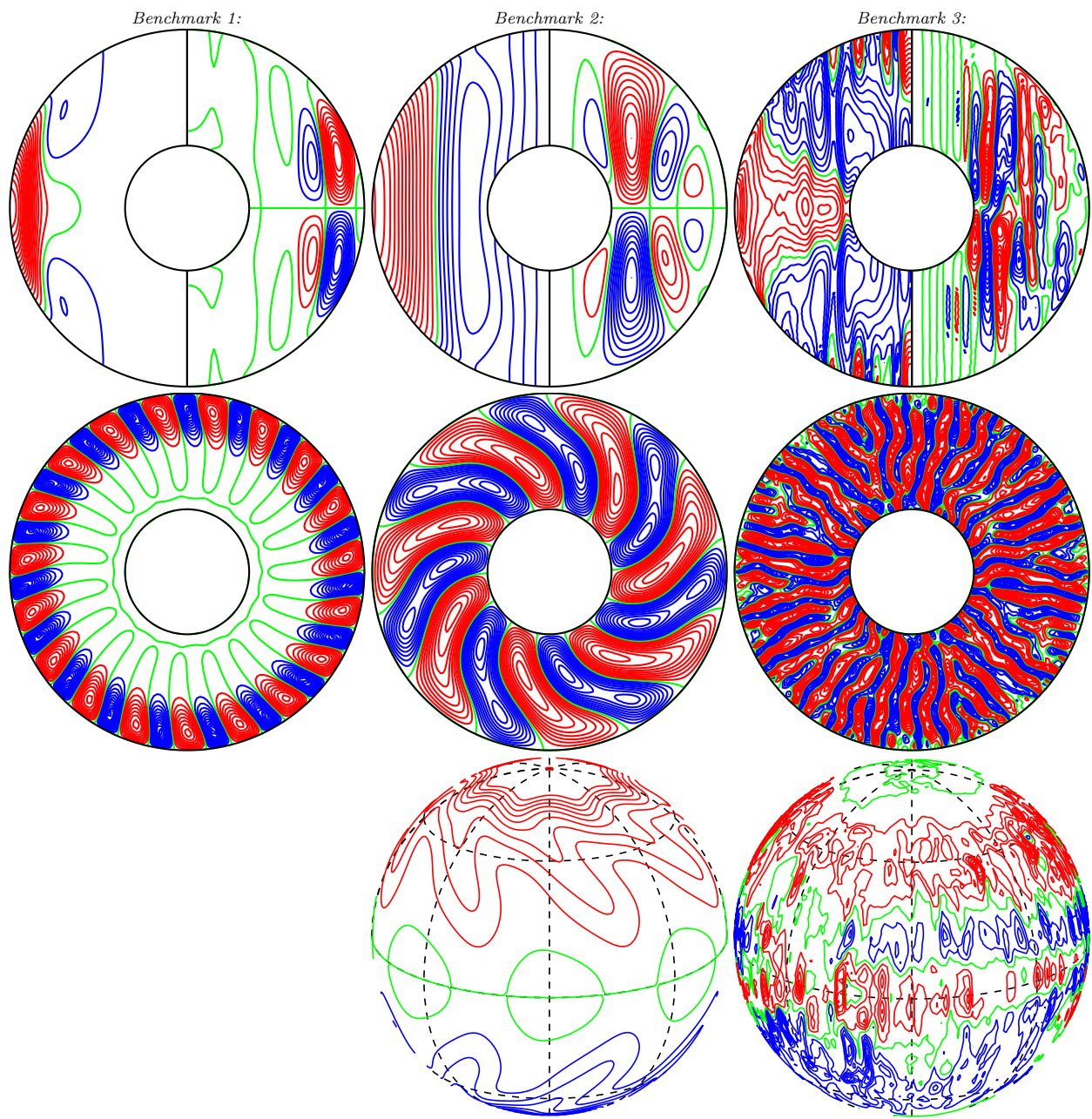,width=\columnwidth,clip=}
\caption{Solutions structures of benchmark cases 1, 2 and 3 (left to
right). The first plot in each column shows azimuthally-averaged 
isocontours of $\overline{u}_\varphi$ (left half) and of the
streamlines $r\sin\theta(\partial_\theta \overline{\pol})$ (right
half) in the meridional plane. The second plot in each column shows
isocontours of $u_r$ in the equatorial plane. The third plot in each
column shows isocontours of $B_r$ at $r=r_o$. }
\label{fig01}
\end{figure}

To validate the new code we present a comparison with the anelastic
benchmark simulations recently proposed by \citet{Jones2011}. For the
comparison we employ an alternative parametrization based on the
magnetic rather than  viscous  diffusion time scale, used in the
benchmark models. Our output results from the three benchmark cases  
defined in \citet{Jones2011} are summarized in Table \ref{tab01}, and
selected components of the solution are plotted in Figure \ref{fig01}. 
The mean values and the means of the deviations computed from the
values reported by the four codes participating in \citep{Jones2011}
are also listed in Table \ref{tab01}.
We achieve perfect agreement with the benchmark results for the
hydrodynamic case  and the steady dynamo case, labelled
\textit{Benchmark 1} and \textit{Benchmark 2}. Our results for the
unsteady dynamo case labeled \textit{Benchmark 3} show some
insignificant differences from the values reported in
\citet{Jones2011}. The reason for the discrepancies is the use of
imposed 2-fold azimuthal symmetry and lower resolution in our code to
reduce computing time.   

\section{Differential rotation and dynamo action in the buoyancy-dominated regime}
\label{sec.4}

It is well known, that as buoyancy forcing becomes significantly
larger than the Coriolis force anti-solar differential rotation
develops. Such buoyancy-dominated regime was first identified
by \cite{Gilman1976} and \cite{Gilman1979}, and more recently it was studied by
\cite{Aurnou2007} and \cite{Gastine2013}. More recent studies include
the works of \citep{Guerrero2013a,Guerrero2013} that is closely
tailored to the solar case. These studies consistently found that due
to vigorous mixing angular momentum is homogenized within the whole
volume of the shell, and this leads to a strong retrograde zonal flow
in the equatorial region, and thus to the anti-solar type of rotation
profile. 

\subsection{Transition between rotation-dominated and buoyancy-dominated regimes}
  
\begin{figure*}
\begin{center}
%
\epsfig{file=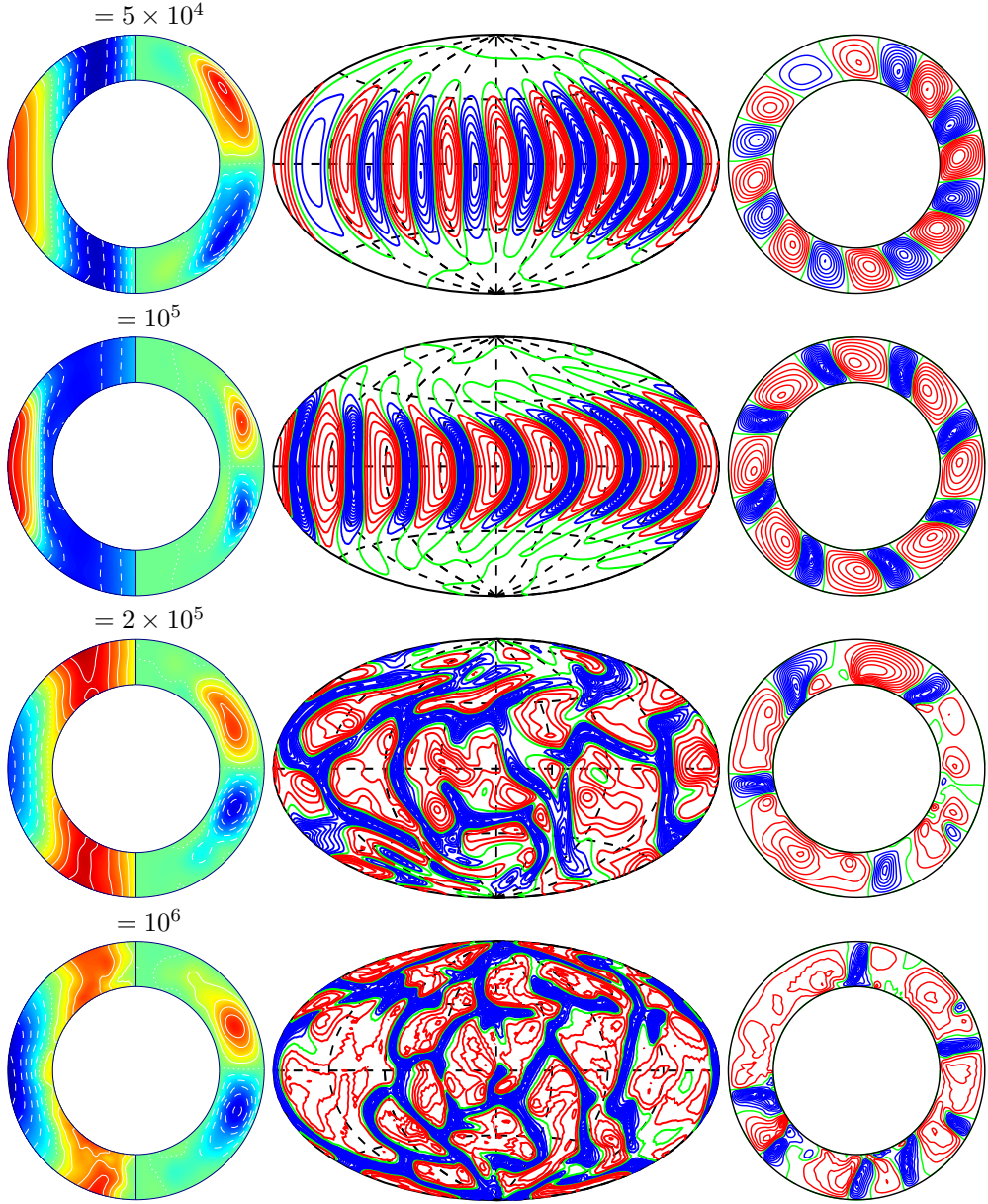,width=0.8\textwidth}
\end{center}
\caption{Structures of convection showing the transition to the
  buoyancy-dominated regime with increasing value of the Rayleigh
  number as indicated in the plot and $\eta=0.65$, $\Pr=0.3$, $\Pm=3$,
  $\tau=200$, $n=2$ and $N_\rho=3$.
The plots in the first column show 
time- and azimuthally-averaged
isocontours of $\overline{u}_\varphi$ (left half) and of the streamlines
$r\sin\theta(\partial_\theta \overline{\pol})$ (right half) in the
meridional plane. The plots in the second column show contours of
instantaneous
$u_r$ on the spherical surface $r=(r_i+r_o)/2$. The plots in the third
column show contours of 
instantaneous
$u_r$ in the equatorial plane. 
}
\label{FIG0010.eps}
\end{figure*}

Here, we investigate the transition from the rotation-dominated regime
to the buoyancy-dominated  regime for non-magnetic and magnetic
(dynamo) cases. 
 Figure \ref{FIG0010.eps} demonstrates the transition from
the solar-type to the anti-solar type of differential rotation in a
set of cases with increasing value of the Rayleigh number $\R$ and all
other parameter values kept fixed. The increase of
the Rayleigh number is the most direct approach to the
buoyancy-dominated regime as $\R$ is a direct measure of the
magnitude of buoyancy forcing. We remark, however, that the
buoyancy-dominated regime can also be reached if other control
parameters are varied. For instance, a decrease in the Coriolis number
$\tau$ at fixed values of the other parameters including $\R$ will equally
well bring convection to the buoyancy-dominated regime since $\tau$ is
a measure of the Coriolis force. Reaching the regime by variation of
the other parameters is also possible but less straightforward.
In the case of Figure \ref{FIG0010.eps} the transition to the buoyancy-dominated
regime happens between $\R=10^5$ and $\R=2\times10^5$. The
Coriolis-dominated regime is characterized by columnar convection
structures that are oriented parallel to the rotation axis and mostly
outside the tangent cylinder, the cylinder that touches the inner core
at the equator. 

In the case of Figure \ref{FIG0010.eps}, columnar
convection has a dominant azimuthal wave number of 8. This number, however,
varies as the parameter values are varied. Since even in the strongly
chaotic regime the azimuthal wave number of convection remains similar to
that near the onset of convection, it is useful to note the study of
\cite{Busse2014} where the critical onset for this problem has
been studied. Differential rotation in the Coriolis-dominated regime
is solar-like and geostrophic, i.e.~constant on cylindrical
surfaces. Meridional circulation is relatively weak but changes from
single to two cells in a hemisphere with the increase of $\R$. In this
connection we note recent observational results by \cite{Zhao2013}
that report two-cell meridional circulation in the solar convection zone.

Convection in the buoyancy-dominated regime for $R> 2\times10^5$ in the
case of Figure \ref{FIG0010.eps} becomes disorganized. Convective
columns are broken, and the convection pattern loses its anisotropy
with respect to the axis of rotation. Convection, for instance, is now not
restricted to the region outside of the tangent cylinder but produces
vigorous flows in the polar regions as well. Meridional circulation
appears to return to a single-cell pattern but the symmetry with
respect to the equatorial plane is lost. This, of course, is due to
the fact that the Coriolis force is no longer dominant and the role of
rotation is much diminished. 

The  most notable effect is the sign reversal of
the differential rotation that switches from the solar-like to
the anti-solar profile. 
\revrs{q4}{The anti-solar differential rotation remains largely constant
  on cylinders parallel to the rotation axis. While some conical
  features are present in the case of Figure \ref{FIG0010.eps} they
  seem to be confined near the surface and appear less significant.
} 

\subsection{Effects of magnetic field on differential rotation}

While solar convection is likely dominated by buoyancy, it is rather
challenging to reconcile the anti-solar differential rotation found in
the buoyancy-dominated regime with observations. It is well
known from helioseismology observations that solar differential
rotation is prograde and strongly non-geostrophic
\citep[e.g.][]{Thompson1996}. \revrs{q2}{In this situation, magnetic effects may
provide one possible mechanism for reversing the anti-solar
differential rotation into the solar-like type \citep{Fan2014,Karak2015,Mabuchi2015}.}
\revrs{q9}{This is not unreasonable to expect. Indeed, it is
well-established that the main effect of self-generated magnetic field
on convection is to impede differential rotation. This effect is
measured by a strong decrease of mean toroidal kinetic energy in
dynamo solutions initially demonstrated by \cite{GroteBusse2001} and
later studied in the parameter space by \cite{Simitev2005a} (see their
Fig.~16) and \cite{Simitev2005b} (see their Fig. 7). It has also been investigated by \cite{AUBERT2005} and \cite{Yadav2013}.}
To explore the effect of magnetic field on buoyancy-dominated
convection as described in the preceding Section, we have performed a
systematic comparison between large sets of simulations of
non-magnetic convection and of self-sustained dynamos.   

\begin{figure}
\begin{center}
\psfrag{up}{\hspace{-6mm}$\overline{u}_\varphi(r_o,\pi/2)$}
\psfrag{R-6}{\hspace{-5mm}$\R\times 10^{-6}$}
\psfrag{R-5}{\hspace{-5mm}$\R\times 10^{-5}$}
\psfrag{E}{$E$}
\psfrag{M}{$M$}
%
\epsfig{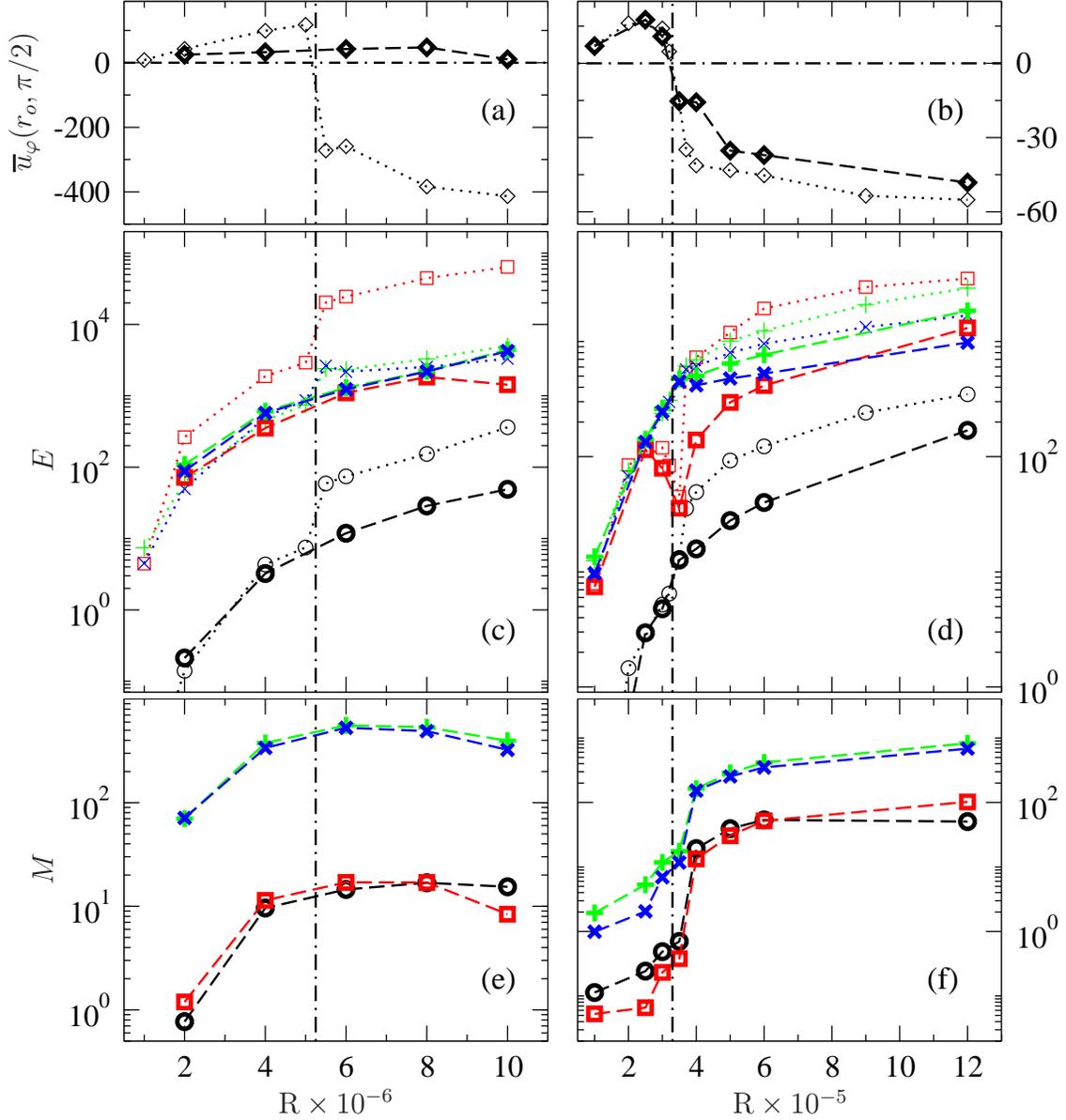}
\end{center}
\vspace*{-6mm}
\caption{
(a,b) Differential rotation at the equator $\overline{u}_\varphi(r_o,\pi/2)$.
(c,d) Average kinetic energy densities and (e,f) average magnetic energy
densities as functions of the Rayleigh number $\R$ in the cases
(a,c,e) for $\eta=0.65$, $\Pr=1$, $\tau=2000$, $n=2$, $N_\rho=3$, and
in the cases (b,d,f) for $\eta=0.65$, $\Pr=0.5$, $\tau=300$, $n=2$, $N_\rho=3$.
Nonmagnetic convection cases are denoted by thin symbols in
(a,b,c,d). Dynamo cases are denoted by thick symbols in all panels and
have $\Pm=2$ in (a,c,e) and $\Pm=6$ in (b,d,f). Black circles, red
squares, green pluses and blue crosses denote $\overline{X}_p$,
$\overline{X}_t$, $\tilde{X}_p$, $\tilde{X}_t$, with $X=E,
M$. Vertical dash-dotted lines denote the transition to buoyancy-dominated regime.
}
\label{cases.engs03.eps}
\end{figure}

\begin{figure}
\begin{center}
\psfrag{Ro,Lo}{\hspace{0mm}{\color{red}$\Ro, \Lo$}}
\psfrag{Rm}{\hspace{0mm}{\color{blue}$\Rm$}}
\psfrag{R-6}{\hspace{-5mm}$\R\times 10^{-6}$}
\psfrag{R-5}{\hspace{-5mm}$\R\times 10^{-5}$}
%
\epsfig{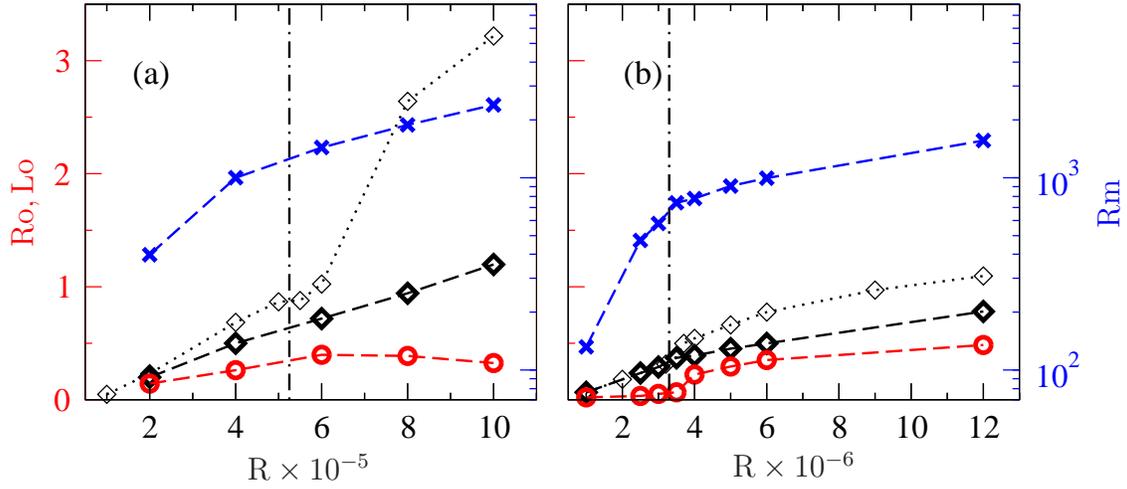}
\end{center}
\vspace*{-6mm}
\caption{
Time-averaged values of the Rossby number Ro (black diamonds) and the Lorentz
number Lo (red circles) measured on the left-hand axis and of the magnetic
Raynolds number Rm (blue crosses) measured on the right-hand
axis. Panel (a) shows the same sequences illustrated in Figure
\ref{cases.engs03.eps}(a,c,d) and panel (b) shows the same sequences
illustrated in Figure \ref{cases.engs03.eps}(b,d,e). The thin symbols
indicate the values of Ro for non-magnetic convection and bold symbols
represent the dynamo cases.
}
\label{FIG0021.eps}
\end{figure}

Figures \ref{cases.engs03.eps} and \ref{FIG0021.eps} summarize the results of four such
sequences of cases with one pair at Prandtl number $\Pr=1$ and Coriolis number
$\tau=2000$ and the other pair at somewhat smaller values $\Pr=0.5$ and
$\tau=300$. One sequence in each pair is non-magnetic while the other
sequence includes a self-generated magnetic field.  The specific
choice of parameters is motivated by properties of solar convection as
discussed below.  Time-averaged components of the kinetic and magnetic
energy densities  are plotted as a function of the increasing Rayleigh
number $\R$ for both the convection-only cases and the dynamo cases. 
The top row of Figure \ref{cases.engs03.eps} shows the time-averaged
value of the differential rotation at the equator on the outer surface
of the spherical shell, $\overline{u}_\varphi(r_o,\pi/2)$, and serves
as an easily accessible indicator of the type of 
differential rotation. The transition from the rotation-dominated regime
(solar-like rotation) to the buoyancy-dominated regime (anti-solar rotation) is thus easily identified by the sign change of
$\overline{u}_\varphi(r_o,\pi/2)$ (marked by a dash-dotted line in
Figures \ref{cases.engs03.eps} and \ref{FIG0021.eps}). Convection in the absence of magnetic
field is characterized by an abrupt as opposed to a gradual increase
of both differential rotation and meridional circulation.  
The total magnetic energy of the self-sustained field is on average
an order of magnitude smaller than the kinetic energy, but the magnetic
field has a significant effect on both the differential rotation and
meridional circulation. In contrast to the non-magnetic case, they are
strongly reduced and show no abrupt change in their values. 

In the sequences shown in Figure \ref{cases.engs03.eps}(a,c,e) the transition from the solar-like to
anti-solar differential rotation is, in fact, suppressed. 
While we expect that if buoyancy is further increased, i.e.~by
increasing $\R$, convection will eventually arrive once again at the
transition to prograde rotation, we emphasize that the suppression of
the transition happens over a large interval of 
$\R$ comparable to the interval between the onset of convection and
the solar-antisolar transition itself.

In the sequences shown in \revrs{q10}{Figure \ref{cases.engs03.eps}(b,d,f)} 
the suppression of the solar-antisolar transition is not observed, even
though the effects of decreasing differential rotation and meridional
circulation are visible. This difference illustrates other important
parameter dependences, primarily those on the Prandtl and the
Coriolis numbers, $\Pr$ and $\tau$, respectively. In the region of low
values of Prandtl and Coriolis numbers convection is known to be rather
different from columnar convection in that it takes the form of
an equatorial belt of large cells attached near the outer
surface of the spherical shell
\citep{Ardes1997,Simitev2004}. Differential rotation generated by
equatorially-attached convection is typically less affected by the
braking effect of the magnetic field \citep{Simitev2005a}.

\revrs{q11}{Figure \ref{FIG0021.eps} shows the values of the Rossby,
Lorentz and magnetic Reynolds number in dependence of the Rayleigh
number of the same sequences as shown in Figure
\ref{cases.engs03.eps}. The transition from rotation-dominated to
buoyancy-dominated convection happens at about $\Ro=1$ in the sequence
of  Figure \ref{cases.engs03.eps}(a,c,e) and at about $\Ro=0.5$ in the
sequence of  Figure \ref{cases.engs03.eps}(b,d,f). These values are 
similar to the ones reported by \cite{Gastine2014}, \cite{Karak2015} and
\cite{Mabuchi2015} but a weak dependence on the Prandtl and the
Coriolis numbers appears to exist. The dynamo effects do not appear to
affect the value of the Rossby number for transition in agreement with
\cite{Karak2015} and \cite{Mabuchi2015}.}

Further notable effects of the magnetic field are summarized in Table
\ref{table02}. These effects are illustrated in terms of one selected
strongly chaotic case discussed below. 

\subsection{Structure and dynamics of convective flows and magnetic fields in the buoyancy-dominated regime}

\begin{figure}
\begin{center}
%
\epsfig{file=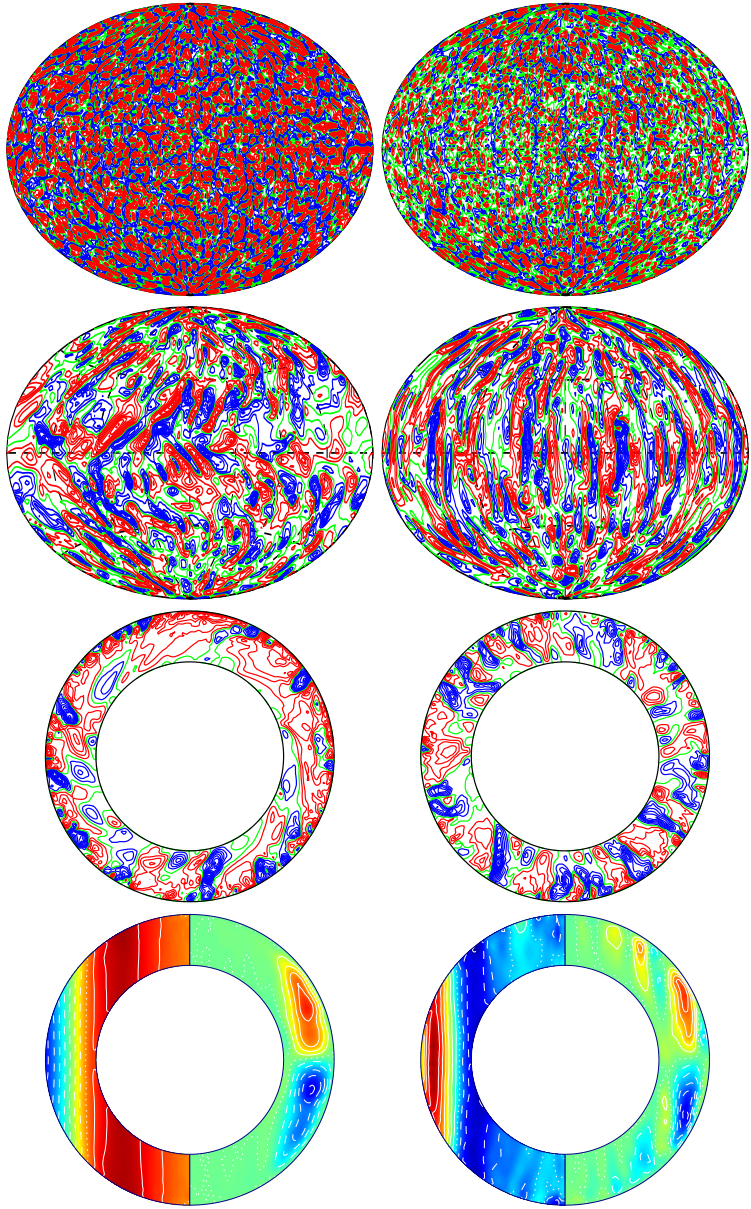,width=0.6\columnwidth,clip=}
\end{center}
\caption{Comparison between non-magnetic convection (left) and dynamo
(right) at identical parameter values
$\eta=0.65$, $\Pr=1$, $\tau=2000$, $\R=10^7$, $n=2$, $N_\rho=3$ and for
  the dynamo
$\Pm=2$.
Plots in the first row show $u_r$ at $r=0.95+r_i$, plots in the second row
show $u_r$ at $r=0.3+r_i$, plots in the third row show $u_r$ in the equatorial
plane, plots in the fourth row show isocontours of the differential rotation
$\overline{u}_\varphi$ (left half) and of the streamlines
$r\sin\theta(\partial_\theta \overline{\pol})$ (right half) in the
meridional plane.
The plots in the first three rows are instantaneous snapshots, while
the density plots in the fourth row are time-averaged. 
}
\label{FIG0030.eps}
\end{figure}

Figure \ref{FIG0030.eps} shows a comparison of the spatial structures
of the convective flow of a non-magnetic case and those of a
self-sustained magnetic dynamo case at identical parameter values
$\eta=0.65$, $\Pr=1$, $\tau=2000$, $\R=10^7$, $n=2$, $N_\rho=3$ and for
the dynamo solution $\Pm=2$. At $\eta=0.65$ the shell thickness is
slightly thicker than the thickness of the solar convection zone and
is selected for ease of numerical simulation. The typical size of
convective structures is related to the thickness of the shell and
thus thinner shells require spherical harmonics decomposition of
higher order and degree to resolve the angular structure of the flow. 
The value $Pr=1$ is appropriate in the sense that
turbulent mixing tends to homogenize the flow and molecular
diffusivities are replaced by effective turbulent diffusivities of
similar magnitude. At $\tau=2000$ the Coriolis number is moderately
but not excessively large reflecting the model assumption that
the flow in the deep convection zone is buoyancy rather
than rotation dominated. The values of the polytropic index $n=2$
is adequate, while the value of the density scale height $N_\rho=3$ is
much lower than estimated for the solar convection zone. However,
increasing, $N_\rho$ much beyond 5 becomes computationally very
demanding. Finally, the value of the Rayleigh number has been selected
such that the non-magnetic convection case is located in the
buoyancy-dominated regime, the onset of which is at $\R=5.2\times10^6$ for these
parameter values.

In the case of non-magnetic convection shown in the right column of 
Figure \ref{FIG0030.eps} differential rotation is in 
anti-solar direction. It is strongly geostrophic i.e.\ constant on
cylinders parallel to the rotation axis. Differential rotation is 
monotonously increasing towards the outer surface of the spherical
shell. Note that this is also true for solar-like differential rotation
in the rotation dominated regime. 
The structure of the flow changes significantly with radius. The first
two rows of Figure \ref{FIG0030.eps} show isocontours of the radial velocity
near the surface and somewhat below mid depth within the spherical
shell. The flow near the surface is a patchwork of small-scaled up-
and down-wellings distributed in a very chaotic pattern over the full
surface of the spherical shell. No visible structure can be discerned
and the location of the convective cells changes chaotically in time
(not shown). The scale of the convective structures increases in depth
and near the equator elongated convective cells tilted clockwise in
the northern hemisphere and tilted anticlockwise in the southern
hemisphere are found. On average this equatorial pattern drifts in the
retrograde direction carried away by the strong anti-solar
differential rotation. The so described radial structure is also
evident in the contour lines of the radial velocity in the equatorial
plane plotted in the third row of \ref{FIG0030.eps}. Finally, the
meridional flow takes the form of two large circulations in poleward
direction at the surface of the shell. The meridional circulations are
nearly strictly mirror-symmetric with respect to the equatorial
plane. This and the symmetry of the differential rotation are
remarkable large-scale coherent features of this otherwise very
chaotic solution.

As noted above the magnetic energy is significantly lower than the
kinetic energy of the flow.  The influence of the magnetic
field on convection, however, in the dynamo case shown in the left column of
Figure \ref{FIG0030.eps} is quite remarkable.  The most notable effect is, of
course the reversal of the direction of differential rotation
from the anti-solar to solar-like type. A further remarkable difference
is that the maximum of the differential rotation occurs
in the depth of the spherical shell rather than at the surface as is
always observed in the case of non-magnetic convection.  This is
potentially a significant effect as it means that there is a
negative gradient of differential rotation in the subsurface layer of
the shell. 

In the magnetic case the structure of the flow also changes
significantly with radius. The first two rows of Figure \ref{FIG0030.eps}
show isocontours of the radial velocity at the same radial values as
in the non-magnetic case. The flow near the surface is a again a
patchwork of small-scaled up- and down-wellings distributed in a very
chaotic pattern. No visible structure can be discerned
and the location of the convective cells changes chaotically in time
(not shown).  An important effect of the magnetic field is that in the magnetic dynamo case 
convection appears to be stronger in the polar regions rather than
outside the tangent cylinder. The main difference, however, appears in
depth. Large scale convective columns arranged in a cartridge belt
pattern within the tangent cylinder and spanning both hemispheres to
about mid latitudes are clearly visible. The columns drift in the
prograde direction due to the solar-like differential rotation.  This
is also a significant observation because it indicates that very
little may be inferred for the structure of deep convection from
observations of near surface flows. In particular, it is not known
whether large scale convective columns exist in the deep solar
convection zone or not. Our results indicate large scale convective
columns hidden from view by much smaller-scale chaotic convection with
no discernible structure is a likely dynamical possibility. 
Finally, the meridional flow of the magnetic dynamo case appears
rather disorganized with a number of smaller scale circulations
appearing in both hemispheres as shown in the bottom row of Figure
\ref{FIG0030.eps}.

\begin{figure}
\begin{center}
%
\epsfig{file=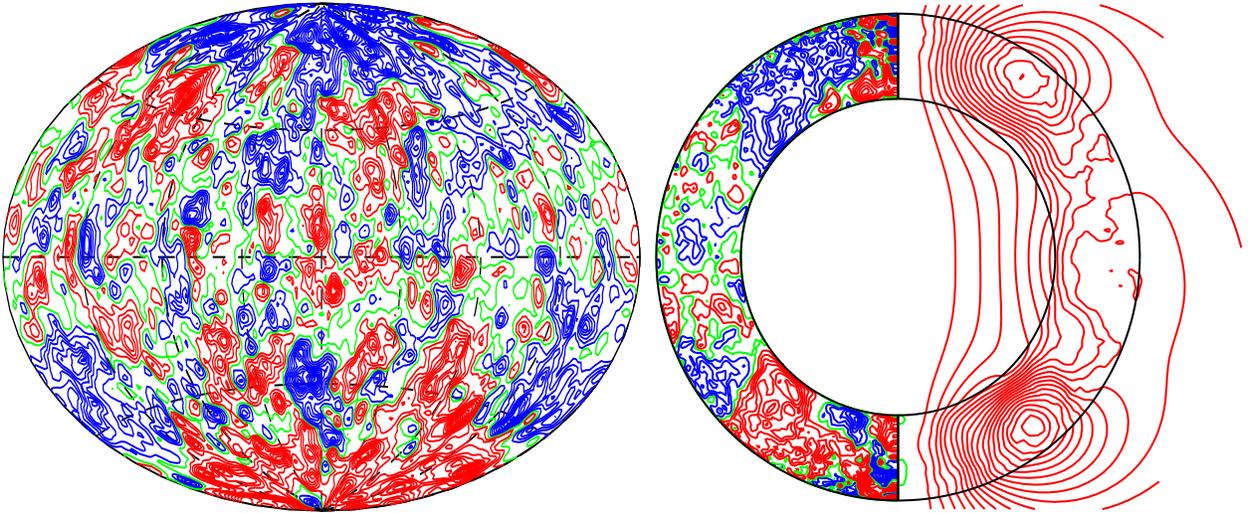,width=\columnwidth,clip=}
\end{center}
\caption{
Magnetic field components of the same case as in the right column of Figure
\ref{FIG0030.eps}.
The left plot shows contours of  $B_r$ at $r=1.13+r_i$, the right plot
shows contours $\overline{B}_\varphi$ and meridional field lines.
}
\label{FIG0040.eps}
\end{figure}

The structure of the generated magnetic field in the dynamo case is
shown in Figure \ref{FIG0040.eps}. The magnetic field has a
large-scale dipole component emerging from a patchwork of small scale
magnetic features. The dipole is mainly supported by strong polar
magnetic flux tubes in the polar region, which in turn are due to the
relatively strong polar convection. The predominant polarity is less
clear in the equatorial region where the magnetic field structures are
smaller in scale and of both polarities. The dipole solution is
non-oscillating. Unfortunately, we have not been able to locate
oscillating dynamos in this regime and to observe the direction of dynamo  
wave propagation. This is left for future studies. 

\revrsa{\section{Remarks on anelastic dynamos in the rotation-dominated regime}}
\label{sec.49}
\revrs{q4}{\revrs{q12}{\revrs{q13}{
While the attention in this paper is focussed on dynamo effects near
the transition from rotation-dominated to buoyancy-dominated
convection, in this section we wish to demonstrate that conical
profiles of differential rotation as well as regular and persistent
dynamo oscillations can be found in the parameter space of 
our minimal convection-driven dynamo model without recourse to
additional modelling assumptions. We also take the opportunity to
elucidate some of the points made about buoyancy-dominated
dynamos above by comparison with features in the rotation-dominated regime.} }}

\begin{figure}
\begin{center}
\epsfig{file=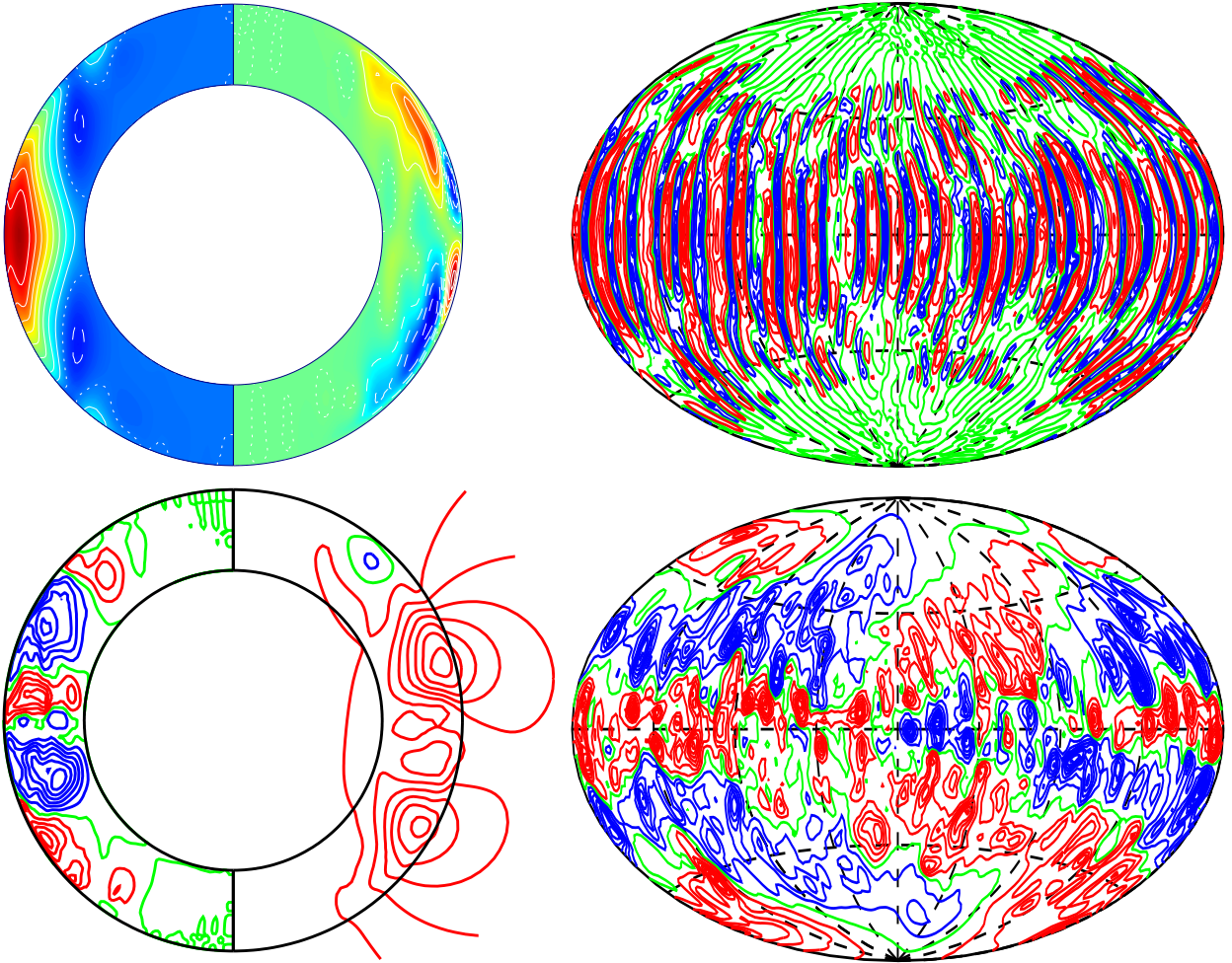,width=\textwidth,clip=}
\end{center}
\caption{
Flow and field structures in the case $\eta=0.65$, $\Pr=1$,
$\tau=2\times10^3$, $\R=1.8\times10^6$, $\Pm=8$, $n=2$, $N_\rho=3$
with no-slip condition on the inner boundary.
The top left plot shows time-averaged isocontours of the differential 
      rotation $\overline{u}_\varphi$ (left half) and of the
      streamlines $r\sin\theta(\partial_\theta \overline{\pol})$
      (right half) in the meridional plane. The top right plot shows contours of instantaneous
      $u_r$ at $r=0.5+r_i$. The bottom left plot shows contours of
      instantaneous $\overline{B}_\varphi$ (left half) and meridional
      field lines (right half). The bottom right plot shows contours
      of instantaneous  $B_r$ at $r=1.13+r_i$.
}
\label{FIG0050.eps}
\end{figure}

\revrs{q4}{
Perhaps the only clear example of a conical profile of differential rotation
in a single-layer simulation with spherically-symmetric boundary
conditions is that reported in the work of \cite{Brun2002}. However, these authors
use subgridscale parametrisation of diffusivities which clearly
affects results. Following  \cite{Miesch2006}, the majority of models that report conical profiles
seem to impose a non-zero latitudinal gradient of entropy as their
bottom boundary condition \citep[e.g.][]{Fan2014} or to 
include anisotropic heat conductivity \citep[e.g.][]{Karak2015} or a stably
stratified layer at the bottom of the convection zone
\citep[e.g.][]{Mabuchi2015} all of which increase the
baroclinicity and induce conical profiles.} \revrs{q4}{In this
  context, Figure \ref{FIG0050.eps} shows 
an example of differential rotation with 
some conical
features in the lower part of the convection zone, obtained in our minimal self-consistent formulation of the
problem. The parameter values of this run are the same as those for the sequence
of cases reported in Figure \ref{cases.engs03.eps}(a,c,e), with a
value of the Rayleigh number that places it in the rotation-dominated
regime, and a somewhat larger value of $\Pm$ which is known to promode
stronger dipolar fields \citep{Simitev2005a}. This run differs from
the latter sequence only in that it uses a 
no-slip velocity condition on the inner spherical boundary. Convection
in the polar regions is weak if not fully absent.}
\revrs{q13}{Convective flows are
confined outside of the tangent cylinder and this is where the dynamo
process is also located resulting in a magnetic field that is strong near the
equator and at midlatitudes but weak in the polar regions. We wish to
contrast this situation with the situation discussed in connection to
the dynamo case presented in Figures \ref{FIG0030.eps} and
\ref{FIG0040.eps}. This comparison makes it rather obvious that in the
latter case vigorous convection in the polar regions gives rise to
strong magnetic field in the same 
regions.} \revrs{q13}{The dynamo shown in Figure \ref{FIG0050.eps} is an
oscillatory dynamo and the comparison with the case of
Figure \ref{FIG0040.eps} elucidates the reasons why the latter is
non-oscillatory. While in the case of Figure \ref{FIG0050.eps} both the
magnetic field and the differential rotation achieve
their maxima in the same region (the equatorial region), in the case
of Figure \ref{FIG0040.eps} the regions where the maximal amplitude of
the magnetic field and of the differential rotation occur do not coincide
which is detrimental to $\alpha\Omega$ oscillations \citep{Simitev2006b,Warnecke2014}.
}

\begin{figure}
\begin{center}
\epsfig{file=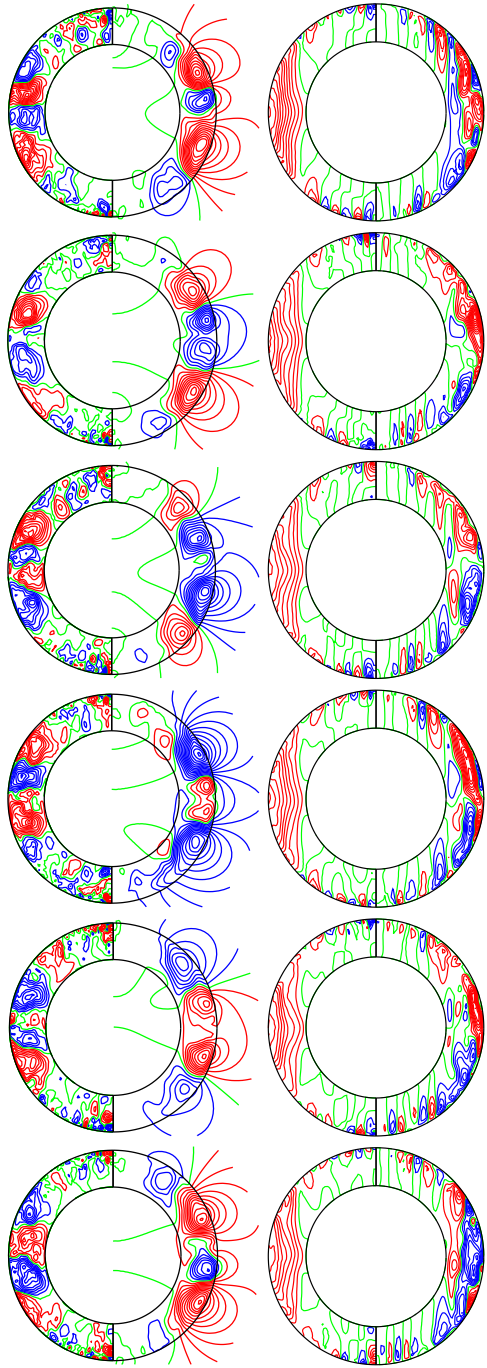,height=0.85\textheight,clip=}
\end{center}
\caption{
One period of dipolar oscillations in the case $\eta=0.65$, $\Pr=1$,
$\tau=2\times10^3$, $\R=2.5\times10^6$, $\Pm=4$, $n=2$, $N_\rho=3$ with
no-slip condition on the inner boundary. Time between plots is $\Delta
t=0.09$ staring at $t_0=96.8174$ in the time series shown in figure
\ref{FIG0061.eps}. 
The first row shows contours $\overline{B}_\varphi$ to the left and
meridional field lines to the right. 
The second row shows contours
$\overline{u}_\varphi$ to the left and the streamlines
$r\sin\theta(\partial_\theta \overline{\pol})$ to the right all in the
meridional plane.
}
\label{FIG0060.eps}
\end{figure}

\begin{figure}
\psfrag{H012}{$H_0^{1,2}$}
\psfrag{G012}{$G_0^{1,2}$}
\begin{center}
\epsfig{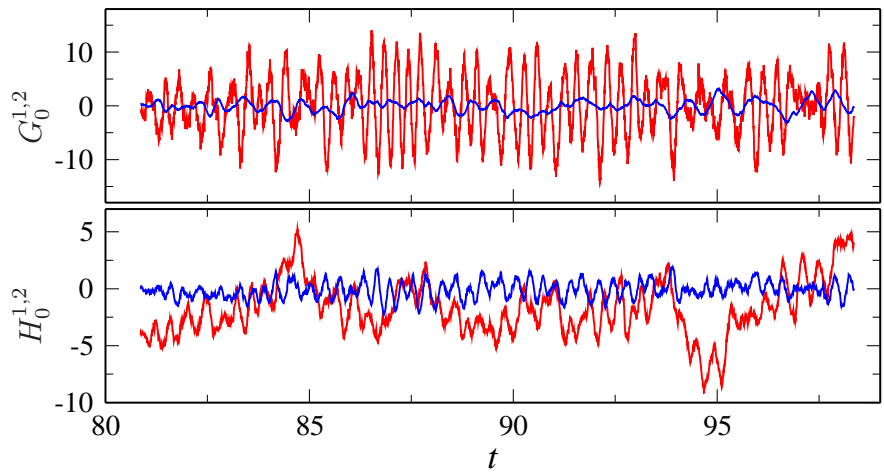}
\end{center}
\caption{
Regular and persistent dipole-dominated oscillations shown in the time
series of the axisymmetric toroidal coefficients $G_0^{1}$ (red) and $G_0^{2}$
(blue) and the axisymmetric poloidal coefficients $H_0^{1}$ (red) and
$H_0^{2}$ (blue) describing the main dipolar and quadrupolar
contributions in the spherical harmonic expansion of the magnetic
field in the case shown in figure \ref{FIG0060.eps}. 
}
\label{FIG0061.eps}
\end{figure}

\revrs{q12}{To illustrate the oscillations in question, we present in
Figure \ref{FIG0060.eps} one period of a predominantly dipolar dynamo
wave. The parameter values of this run are identical to the cases
shown in Figure \ref{FIG0050.eps} except for a slightly larger value of the
Rayleigh number which helps to make the oscillations more regular. 
The dynamo wave is driven by the $\alpha\Omega$ mechanism first
proposed by \cite{Parker1955} and later confirmed in three-dimensional
simulations by \cite{Simitev2006b}, see also \citep{Simitev2012a,Schrinner2012,Warnecke2014}.
The dynamo wave propagates in the direction of the poles. 
This case shows similar conical features of the differential rotation 
profile, and Figure \ref{FIG0060.eps} also illustrates their
variations in time. 
Figure \ref{FIG0061.eps}
shows the dominant dipolar and quadrupolar components
of the magnetic field represented by time series of the appropriate
coefficients in the spherical harmonic expansions of the toroidal and
poloidal scalars of the magnetic field. The time series show that the
oscillations are very regular and persistent over the course of the
simulation and that the dipole is dominant.
}

\revrsa{We wish to conclude this section with the remark that the
parameter space of the minimal model formulation seems to merit 
further investigation especially in the case when self-generated
magnetic fields are present.}

\section{Conclusion}
\label{sec.5}

We have presented in this paper several sets of convective dynamo simulations in
\revrsa{density-stratified} rotating spherical fluid shells based on a
\revrsa{physically consistent anelastic model with a minimum number of 
parameters.}
The computations are performed using a new simulation code that 
is also presented here for the first time along with code validation
results against published benchmark solutions.
\revrsa{We demonstrate that conical differential rotation
profiles and persistent regular dynamo oscillations can be obtained in
the parameter space of this minimal formulation without recourse to
additional modelling assumptions.}
\revrsa{The main focus of the work is placed on extending the dynamo simulations}
into a ``buoyancy-dominated'' regime where the 
buoyancy forcing is dominant while the Coriolis force is no longer
balanced by pressure gradients and where strong anti-solar
differential rotation develops as a result. The dynamo solutions are
compared to identical sets of non-magnetic convection solutions
\revrsa{to reveal the effects of the self-sustained magnetic field on
 convection in general and on the differential rotation in particular.}
The most significant results are summarized in Table \ref{table02} and
below we discuss some similarities between our solutions and solar and
stellar convection. 

\revrs{q2}{\revrs{q3}{We also wish to compare our results to
studies on a similar topic reported in the recent literature, in
particular the works of \cite{Fan2014}, \cite{Karak2015} and
\cite{Mabuchi2015} where dynamo simulations are reported and of
\cite{Gastine2013} and \cite{Hotta2015} where hydrodynamic simulations
are reported. Before we comment on similarities and differences in
solutions, we wish to point out that with the exception of the model
considered by \cite{Gastine2013}, the models considered by the other
groups have significant and essential differences compared to
ours. The models of \cite{Hotta2015}, \cite{Karak2015} and
\cite{Mabuchi2015} are fully compressible models in contrast to our
anelastic approximation. \cite{Hotta2015} use artificially enhanced
viscosity, a radius dependent cooling term and are interested mainly
in the properties of the near-surface convection layer. The model of
\cite{Mabuchi2015} consists of two layers -- a stably stratified layer
surrounded by a convective envelope. The models of \cite{Fan2014} and
\cite{Karak2015} consider wedges, i.e. partial spherical shells, and
are not fully spherical, so polar convection is effectively not
represented in their solutions. This partly explains why they find
oscillatory dynamos. Conical differential rotation profiles are
promoted in the latter four models due to the inclusion of secondary
physical effects: \cite{Fan2014} impose a latitudinal entropy gradient
as a bottom boundary condition, \cite{Fan2014}, \cite{Hotta2015} and
\cite{Karak2015} include radial variation in diffusivities and
parametrisations of 
unresolved scales. Numerical implementations also differ -- all codes
except that of \cite{Gastine2013} are based on finite-difference
methods while ours is pseudo-spectral. An artificial radial magnetic
field condition is imposed on the outer boundary in these studies
which may significantly distort dynamo effects. Despite the differences in
modelling strategy, it is significant that we find a number of
similarities in our simulations.  This increases the confidence in the
robustness of the reported results.}} 

Oscillations obtained in dynamos generated within the 
rotation-dominated regime \revrs{q12}{with few exceptions
\citep{Warnecke2014}, appear to always travel in the poleward
direction much like illustrated in Figure \ref{FIG0060.eps}. For
references see} \citep{Simitev2006b,Simitev2012a}, \revrs{q12}{and also
\citep{Schrinner2012}. There is some evidence that dynamo waves can
travel towards the equator when a negative radial gradient of the 
differential rotation profile exists \citep{Simitev2012b}, and when
in addition to the latter the $\alpha$-effect, proportional to
$-(\nabla \times \u) \cdot \u + (\nabla\times\B)\cdot \B/\rhobar$, is 
positive (negative) in the northern (southern) hemisphere
\citep{Warnecke2014}.}  This evidence supports an early analysis by \cite{Yoshimura1975}. 
\revrs{q2}{While we commonly find dipolar oscillations in the rotation-dominated
regime, our dynamos in the buoyancy-dominated regime do not
oscillate. \revrs{q13}{This is in contrast to the results of
\cite{Fan2014} and \cite{Karak2015}  which may be attributed to the
absence of polar convection in their simulations due to the absence of conical polar sections
 used in their geometrical configuration. On the other hand, we agree with
\cite{Mabuchi2015} who use a full spherical shell and find that dynamos with anti-solar rotation are
predominantly dipolar  and non-oscillatory.}} 
Despite being relatively weak the self-sustained magnetic fields in the
buoyancy-dominated regime reported in the present study are able to
reverse the direction of differential rotation from anti-solar to 
solar-like. \revrsa{From the perspective of oscillations, it is
significant that} we find that differential rotation attains a maximum
inside the shell and \revrsa{that a negative radial gradient is
persistently  maintained in the near-surface layer}. This may facilitate equatorward dynamo
wave propagation in the buoyancy-dominated regime. \revrs{q2}{The differential
rotation in our buoyancy-dominated dynamos, e.g.~Figure \ref{FIG0030.eps}, has
a cylindrical profile, in contrast to the more conical profiles reported
by \cite{Fan2014} \cite{Karak2015} and \cite{Mabuchi2015}. This
difference is almost certainly caused by the fact that a non-zero latitudinal
gradient of entropy is imposed as a bottom boundary condition by
\cite{Fan2014}, that an anisotropic heat conductivity is used by
\cite{Karak2015}, and that a stably stratified layer at the
bottom of the convection zone is present in
\citep{Mabuchi2015}.}

We find that the dynamo-generated magnetic field can suppress the
transition from the solar-like to the antisolar-like rotation
profile \revrs{q2}{thus confirming similar findings reported by
  \cite{Fan2014} \cite{Karak2015} and \cite{Mabuchi2015}.}  In this
case the
convection is significantly stronger near the poles than in the
equatorial region, \revrsa{leading to a predominantly dipolar dynamo
where both the toroidal and the poloidal fluxes are stronger in
the polar regions compared to equatorial regions.} Such dynamo regime
with concentration of magnetic
field in the polar regions may explain the observations of polar
starspots in young solar-type stars which exhibit reduced but still
the solar-type differential rotation
\citep{Brown2014}, also see \citep{Yadav2015}. \revrs{q13}{While starspots reduce locally the
vigour of convection in the near surface layer of the convection zone,
sufficiently strong convection in depth is required to generate
magnetic fields that are large enough to cause starspots in the first
place. Our calculations do not have sufficient resolution to resolve
strongly turbulent stellar near surface layers.}

\revrs{q3}{Our simulations confirm the findings of \cite{Gastine2013}
and \cite{Hotta2015}} that different regimes of convection occur in the
inner and at the outer part of the spherical shell simultaneously such
that organized geostrophic convection columns are hidden below a
near-surface layer of well-mixed highly-chaotic
convection. \revrs{q3}{Both of the latter studies are non-magnetic and
the work of \cite{Hotta2015} reports simulations in the
rotation-dominated regime only. The model of \cite{Gastine2013} is
rather similar to ours and it is one of the aims of the present paper to extend
their analysis through considerations of dynamo effects. 
On the Sun small scale turbulent convection is clearly observable in the
subsurface layer of the solar convection zone while simulations
inevitably find some columnar structures.
Evidence of different convection morphology as a function of radius is significant because
it provides a bridge between observations and simulations.} 
The deeper large-scale organized convection columns are likely to play
important role in the solar dynamo and its magnetic cycles.

\section*{Acknowledgment}
This research has been supported by the NASA Grants NNX14AB70G and
NNX09AJ85G and by the Leverhulme Trust Research Project Grant RPG-2012-600.
The hospitality of Stanford University, UCLA and NASA Ames Research
Center is gratefully acknowledged. RDS enjoyed a period of study
leave granted by the University of Glasgow.


\newpage

\vtwors{
\begin{appendix}
\section{Spectral projection of the toroidal-poloidal governing equations}
\label{appendix}

Scalar equations for $v$ and $w$ are obtained, and effective pressure 
gradients are eliminated by taking $\ru\cdot\n\x\n\x$ and
$\ru\cdot\n\x$ of Equation~\eqref{gov.02}. Similarly, equations for
$h$ and $g$ are obtained by taking $\ru\cdot \n\x$ and $\ru\cdot$ of
Equation~\eqref{gov.04}. The scalar unknowns $\pol$, $\tor$, $\polB$,
$\torB$ and $S$  are then expanded in Chebychev
polynomials $T_p$ in the radial direction $r$, and in spherical
harmonics $Y_l^m$ in the angular directions $(\theta,\varphi)$ as
shown in Equation \eqref{decomp}. After a standard Galerkin projection
procedure in the angular directions $(\theta,\varphi)$ the following
set of partial differential equations for the spectral expansion
coefficients $\{\pol_l^m(r,t), \tor_l^m(r,t), \polB_l^m(r,t),
\torB_l^m(r,t), S_l^m(r,t)\; :\; l=1\ldots\Lmax,\, m=-l\ldots l\}$ is
obtained 
{\small
\bs
\label{pol-tor-eqns-lm}
\begin{gather}
\label{tor-eq-lm}
\hspace*{-75mm}
\dd{t}{\tor_l^m} - 
\left(\partial_r^2+\f{4}{r}\dd{r}{}+\f{2-l(l+1)}{r^2}\right)\tor_l^m 
=
- \zeta^n \Xi_2 \tor_l^m\\
\hspace*{5mm}
-\f{1}{l(l+1)}\zeta^{n}\left[ \ru\.\n\x 
\Bigg(
\Xi_1 \dd{r} \m
 + \tau \Big(\hat{\v k} \x  \f{\m}{\zeta^{n}}\Big)
 +\left(\Big(\n \x \f{\m}{\zeta^{n}}\Big)\x \f{\m}{\zeta^{n}} \right)
 -\f{1}{\zeta^n} (\n\x\B)\x\B
\Bigg)\right]_l^m,
\nonumber\\
\label{pol-eq-lm}
\hspace*{-30mm}
\dd{t}{} \Dl \pol_l^m
- \Dl \Ml \pol_l^m 
= 
- \El \pol_l^m
- \f{1}{r^2} \Xi_3 l(l+1) \pol_l^m
-\f{2}{3} n^2 \f{(\zeta')^2}{\zeta^{n+2}}\f{l(l+1)}{r^2}\pol
- \f{\R}{\Pr} \f{1}{r^3} S_l^m
\\
\hspace*{10mm}
+\f{r}{l(l+1)} \left[\ru\.\n\x\n\x 
\Bigg(
\Xi_1 \dd{r} \m
 +\tau \Big(\hat{\v k} \x  \f{\m}{\zeta^{n}}\Big)
 +\left(\Big(\n \x \f{\m}{\zeta^{n}}\Big)\x \f{\m}{\zeta^{n}} \right)
-\f{1}{\zeta^n} (\n\x\B)\x\B
\Bigg)
\right]_l^m,
\nonumber \\
%
%
\label{entropy-eq-lm}
\hspace*{-68mm}
\dd{t}{}S_l^m
-\f{1}{\Pr}\Ml S_l^m
= 
\f{1}{\Pr}(n+1)\f{\zeta'}{\zeta}\dd{r}{} S_l^m
-\left[\f{1}{\zeta^{n}} \n\. (S\m)\right]_l^m 
\\ \nonumber 
\hspace{30mm}
+\left[\f{c_1 \Pr}{\R\zeta}\Big( Q_v + \f{1}{\Pr_m\zeta^n} (\n\x\B)^2\Big)\right]_l^m
- \f{l(l+1)}{r} \f{\pol_l^m}{\zeta^{n}}
  \f{1}{r^2}\f{1}{\zeta^{n+1}}
 \f{c_1 n d\zeta(r_i)^n\zeta(r_o)^n}{\zeta(r_o)^n-\zeta(r_i)^n},\\
%
%
\dd{t}{} \polB_l^m
-\f{1}{\Pm}\left(\partial_r^2 -\f{l(l+1)}{r^2}\right) \polB_l^m
=
\f{r^2}{l(l+1)} \left[\ru\.\n\x\Big(\f{\m}{\zeta^{n}}\x\B \Big)\right]_l^m,\\
\dd{t}{} \torB_l^m
-\f{1}{\Pm}\left(\partial_r^2 -\f{l(l+1)}{r^2}\right) \torB_l^m
=
\f{r^2}{l(l+1)} \left[\ru\.\n\x\n\x\Big(\f{\m}{\zeta^{n}}\x\B \Big)\right]_l^m,
\end{gather}
\es
where the following operators are defined
\begin{gather*}
\Ml = 
\f{1}{r^2} \big(\dd{r}{}r^2\dd{r}{} - l(l+1)\big) 
= 
\partial_r^2 + \f{2}{r}\dd{r}{} - \f{l(l+1)}{r^2},\\
\Dl = \left(\left(\f{1}{\zeta^{n}}\right)'\Big(\dd{r}{}+\f{1}{r}\Big) +\f{1}{\zeta^{n}} \Ml \right), \\
\El = \left(\Xi_2' \Big(\dd{r}{}  +\f{1}{r}\Big) + \Xi_2 \Ml \right),
\end{gather*}
and the following notation of some radial functions is used for brevity
\begin{gather*}
\Xi_1(r)=\f{1}{\zeta^{n}} \f{n \zeta'}{\zeta},\qquad
\Xi_2(r)=\f{1}{\zeta^{n}} \left(\f{3}{r}\f{n \zeta'}{\zeta} +
\left(\f{n
  \zeta'}{\zeta}\right)' \right), \qquad
\Xi_3(r)=\f{1}{\zeta^{n}} \left(\f{1}{r}\f{n \zeta'}{\zeta}-\left(\f{n
  \zeta'}{\zeta}\right)' \right).
\end{gather*}
The expression for the viscous dissipation in Equation
\eqref{entropy-eq-lm} is
\begin{gather*}
Q_v = Q_v^{(1)}+ Q_v^{(2)},\\
Q_v^{(1)} =
2 \left(\dd{r}{\f{v_r}{\zeta^{n}}}\right)^2
  -\f{2}{3}\left(\f{v_r}{\zeta^{n}}\f{n}{\zeta}\dd{r}{\zeta} \right)^2
  + 2\left(\dd{r}{\f{v_r}{\zeta^{n}}} +
  \f{v_r}{\zeta^{n}}\Big(\f{n}{\zeta}\dd{r}{\zeta}+\f{1}{r}\Big)
         +q_v^{(1)} \right)^2
  +2\left(q_v^{(1)}+ \f{1}{r} \f{v_r}{\zeta^{n}} \right)^2,\\
Q_v^{(2)} =
 \left(2\dd{r}{\f{v_\theta}{\zeta^{n}}} - \Big[\n \x
   \f{\m}{\zeta^{n}}\Big]_\varphi \right)^2
+ \left(2\dd{r}{\f{v_\phi}{\zeta^{n}}} + \Big[\n \x
  \f{\m}{\zeta^{n}}\Big]_\theta \right)^2
+ \Big(2 q_v^{(2)} + \Big[\n \x \f{\m}{\zeta^{n}}\Big]_r \Big)^2,\\
q_v^{(1)}=
 \f{\cos\theta}{r\sin\theta} \f{v_\theta}{\zeta^{n}}
 + \f{1}{r\sin\theta}\dd{\varphi}{\f{v_\varphi}{\zeta^{n}}}, \qquad
q_v^{(2)} =
 - \f{\cos\theta}{r\sin\theta} \f{v_\varphi}{\zeta^{n}}
+ \f{1}{r\sin\theta}\dd{\varphi}{\f{v_\theta}{\zeta^{n}}}.
\end{gather*}
Finally, square brackets with a subscript (and superscript) as in
$[\cdot]_\varphi$ or $[\cdot]_l^m$ denote a component of a vector or
an appropriate coefficient in a spherical harmonic expansion, respectively.

Solution of Equations \eqref{pol-tor-eqns-lm} proceeds as described in Section \ref{nummeth}.

}

\end{appendix}
}

\newpage
\begin{table}
\caption{Comparison with the benchmark solutions reported by
  \cite{Jones2011}. \vtwors{The values labelled ``Mean $\pm$ Mean Deviation'' are the
respective means and the means of the deviations computed from the
values reported by the four codes participating in
\citep{Jones2011}.}} 
\renewcommand{\arraystretch}{1}
\small
\begin{tabular}{p{0.2\columnwidth}p{0.19\columnwidth}p{0.19\columnwidth}p{0.19\columnwidth}}
\toprule
& \textit{Benchmark 1:} & \textit{Benchmark 2:} & \textit{Benchmark 3:} \\
& Hydrodynamic  convection & Steady dynamo  & Unsteady dynamo
\\ \midrule
$\eta$    & {0.35} & {0.35} & 0.35\\[-2mm]
$n$       & {2} & {2} & 2 \\[-2mm]
$N_\rho$  & {5} & {3} & 3 \\[-2mm]
$\Pr$       & {1} & {1} & 2 \\[-2mm]
$\Pm$     & {1} & {50} & 2 \\[-2mm]
$\tau$    & {2000} & {1000} & $4\times10^4$\\[-2mm]
$\R$       & {351806} & $8\times10^4$ & $2.5\times 10^7$\\ [2mm]
$N_r$ / $N_r$        & 129 / 129  & 129 / 129   & 111 / 111 \\[-2mm]
$N_l$ / $N_\theta$   & 128 / 128  & 128 / 128   & 120 / 144 \\[-2mm]
$N_m$ / $N_\varphi$ &  129 / 257  & 129 / 257   & 61 / 73 \\[-2mm]
Timestep    & $4\times 10^{-6}$ & $1\times 10^{-6}$  & $1\times 10^{-7}$ \\[2mm]
$E$ & 81.87991 & $4.19405\times 10^5$ & $2.32730\times 10^5$\\[-2mm]
\scriptsize Mean  $\pm$ Mean Dev.  &  \scriptsize $81.680 \pm 0.245$ &\scriptsize $(4.186\pm0.013)\times 10^5$   &\scriptsize  $(2.317\pm0.014)\times 10^5$ \\[-2mm]
$\overline{E}_p$ & 0.02201  & 53.0100 & 100.40\\[-2mm]
\scriptsize Mean  $\pm$ Mean Dev.  &\scriptsize $0.0220\pm0.0001$ &\scriptsize  $52.90\pm 0.15$   &\scriptsize  111.75 $\pm$ 3.75\\[-2mm]
$\overline{E}_t$ & 9.37598 & $6.01725\times 10^4$& $1.81399\times 10^4$\\[-2mm]  
\scriptsize Mean  $\pm$ Mean Dev.  &\scriptsize  $9.3568 \pm 0.0282$
&\scriptsize  $(6.001\pm 0.018)\times 10^4$   &\scriptsize  $(1.355\pm0.008)\times 10^4$\\[2mm]
$M$ & -- & $3.20172\times 10^5$ & $2.58012\times 10^5$ \\[-2mm]
\scriptsize Mean  $\pm$ Mean Dev.  &\scriptsize   &\scriptsize
$(3.194 \pm 0.088) \times 10^5$   &\scriptsize  $(2.413 \pm 0.023)\times 10^5$  \\[-2mm]
$\overline{M}_p$ &  --  & $1.69650\times 10^4$ & $2.91155\times 10^4$ \\[-2mm]
\scriptsize Mean  $\pm$ Mean Dev.  &\scriptsize   &\scriptsize
$(1.692 \pm 0.005)\times 10^4$   &\scriptsize   $(2.155 \pm 0.070) \times 10^4$ \\[-2mm]
$\overline{M}_t$ &  --  & $2.41185\times 10^5$ & $1.17292\times 10^4$ \\[-2mm]
\scriptsize Mean  $\pm$ Mean Dev.  &\scriptsize   &\scriptsize  $(2.412\pm0.028)\times 10^5$   &\scriptsize   $(0.948\pm 0.003)\times 10^4$ \\[2mm]
Luminosity            & 4.19886 & 11.50302 & 42.50992\\[-2mm]
\scriptsize Mean  $\pm$ Mean Dev.  &\scriptsize  $4.19886 \pm 3\times 10^{-6}$ &\scriptsize  $11.503 \pm 4\times 10^{-5}$   &\scriptsize   $42.75 \pm 0.15$  \\
\bottomrule
\end{tabular}
\normalsize
\label{tab01}
\end{table} 

\begin{table*}
\caption{Summary of the effects of self-sustained magnetic field and
  comparison with non-magnetic convection.}
\renewcommand*{\arraystretch}{1.3}
\begin{tabular}{p{0.43\textwidth}p{0.43\textwidth}}
\toprule
\textbf{Non-magnetic convection} & \textbf{Dynamo}\\ \midrule
 Monotonic increase/decrease of differential rotation towards the outer
  surface in solar/antisolar cases.
&
Differential rotation attains a maximum inside shell and a
  subsurface decrease.
\\ \midrule
Retrograde differential rotation in buoyancy-dominated regime.
&
Differential rotation reversed from antisolar to solar-like.
\\ \midrule
No columnar structure at depth.
&
Convective columns visible in depth.
\\
\bottomrule
\end{tabular}
\label{table02}
\end{table*}

\newpage

\newpage

\end{document}